\documentclass[12pt]{article}
\begin{document}
\title{The Mass of the Neutrinos}
\author{B.G. Sidharth\\
International Institute for
Applicable Mathematics \& Information Sciences\\
Hyderabad (India) \& Udine (Italy)\\
B.M. Birla Science Centre, Adarsh Nagar,\\
Hyderabad - 500 063 (India)}
\date{}
\maketitle
\begin{abstract}
In the theory of the Dirac equation and in the standard model, the
neutrino is massless. Both these theories use Lorentz invariance. In
modern approaches however, spacetime is no longer smooth, and this
modifies special relativity. We show how such a modification throws
up the mass of the neutrino at least the electron neutrino to start with.
\end{abstract}
\section{Introduction}
Today we know that neutrinos are described by the two component
equation. Such an equation has a long history. It was proposed by
Hermann Weyl as long ago as 1929. He argued that it would represent
a massless Fermion. However the suggestion was soon rejected because
such a particle would not be invariant under the parity
transformation. Later experiments showed the non conservation of
parity in beta decay, as suggested by Yang and Lee. Salam and Landau
then proposed that neutrinos obey the Weyl equation, discarded
nearly thirty years earlier. The Weyl equation itself is given by
\cite{schweber}
\begin{equation}
\imath \hbar \partial_t \phi = c \vec{\sigma} \, \cdot {\bf p}
\phi\label{e1}
\end{equation}
This equation brings out particles with definite helicity states,
and satisfies the condition for neutrinos. Though the Weyl equation
differs from the four component Dirac equation, it is well known
that the massless Dirac equation too can represent a neutrino,
provided an extra constraint is satisfied. As we will briefly see
below, the solutions of the Dirac equation preserve parity, while
the constraint removes two of the four components of the Dirac
solution, which thus makes the solution non invariant under parity.
We can examine this a little more carefully, by starting with the
Dirac equation for a massless particle
\begin{equation}
\imath \gamma_\mu \partial^{\mu} \psi (x) = 0\label{e2}
\end{equation}
which in Hamiltonian form reads
\begin{equation}
\imath \partial_t \psi (x) = -\imath \gamma^0 \vec{\gamma} \, \cdot
\nabla \psi (x)\label{e3}
\end{equation}
In the usual representation
\begin{equation}
\gamma^0 = \left(\begin{array}{ll} I \quad 0\\ 0 \quad
-I\end{array}\right) \quad \quad \gamma_5 = - \imath
\left(\begin{array}{ll} 0 \quad I\\ I \quad
0\end{array}\right)\label{e4}
\end{equation}
\begin{equation}
\vec{\gamma} = \left(\begin{array}{ll} 0 \quad \vec{\sigma}\\
-\vec{\sigma} \quad 0\end{array}\right)\label{e5}
\end{equation}
We also need
\begin{equation}
\sum = \left(\begin{array}{ll}\vec{\sigma} \quad 0\\
0 \quad \vec{\sigma}\end{array}\right) = \imath \gamma_5 \gamma^0
\vec{\gamma}\label{e6}
\end{equation}
The Hamiltonian can now be written in the form
\begin{equation}
H = \imath \gamma_5 \sum \cdot {\bf p} = \imath \gamma_5 |{\bf
p}|s({\bf p})]\label{e7}
\end{equation}
We can see from (\ref{e7}) that the eigenfunctions of $H$ and
$s({\bf p})$ are eigenfunctions of $\imath \gamma_5$. The four
linearly independent solutions of
$$H u = p_0u$$
with the $z$ axis as the direction of ${\bf p}$ are given by:
\begin{equation}
\quad \quad \left(\begin{array}{ll} 1\\0\\1\\0\end{array}\right)
\quad \quad \left(\begin{array}{ll} 0\\1\\0\\-1\end{array}\right)
\quad \quad \left(\begin{array}{ll} -1\\0\\1\\0\end{array}\right)
\quad \quad \left(\begin{array}{ll}
0\\1\\0\\1\end{array}\right)\label{e8}
\end{equation}
The eigenvalue of $\imath \gamma_5$ for these solutions are
summarized in Table 1.

\begin{table} [tbp]
\caption{} \vspace{10 mm}
\begin{tabular}{|c|c|c|}
\hline $P_0$ & Helicity & Eigenvalue of $\imath \gamma s$\\ \hline
$+$ & $+1$ & $+1$ \\
$+$ & $-1$ & $-1$ \\
$-$ & $+1$ & $-1$ \\
$-$ & $-1$ & $+1$ \\ \hline
\end{tabular}
\end{table}
The first two spinors in (\ref{e8}) represent positive energy
solutions while the last two represent negative energy solutions.
Their helicities are given by respectively $+1, -1, +1$ and $-1$.\\
As can be seen, the eigenvalue of $\imath \gamma_5$ for a positive
energy solution is the same as that of the helicity operator. For a
negative energy solution on the other hand, the eigenvalue of
$\imath \gamma_5$ is opposite that of the helicity operator.\\
As is well known a neutrino is described by a two-component
equation, the plane wave solutions of which have the property that
for $p_0 = +|{\bf p}|$ the helicity is $-1$, and for $p_0 = -|{\bf
p}|$ the helicity is $+1$. For this we require that the plane wave
solutions of (\ref{e3}) need also to satisfy:
\begin{equation}
\psi = -\imath \gamma_5 \psi\label{e9}
\end{equation}
This constraint is invariant for proper Lorentz transformations. To
put it another way, if $\psi$ is a four-component spinor satisfying
(\ref{e3}), the spinor $\psi_n$ defined by
\begin{equation}
\phi = \frac{1}{2} (1 - \imath \gamma_5) \psi\label{e10}
\end{equation}
satisfies the condition (\ref{e9}) If
\begin{equation}
\gamma^0 = \left(\begin{array}{ll}0 \quad I\\
I \quad 0\end{array}\right) \quad \quad \gamma = \left(\begin{array}{ll}0 \quad \vec{\sigma}\\
-\vec{\sigma} \quad 0\end{array}\right)\label{e12}
\end{equation}
\begin{equation}
\gamma_5 = \imath \left(\begin{array}{ll}I \quad 0\\
0 \quad -I\end{array}\right)\label{e13}
\end{equation}
then the spinor $\phi$ is essentially a two-component quantity since
the projection operator $\frac{1}{2} (1 - \imath \gamma_5)$
annihilates the two lower components. The two-component spinor
$\phi$ satisfies the equation
\begin{equation}
- \sigma \cdot {\bf p} \phi  = p_0 \phi\label{e15}
\end{equation}
as we can see if we multiply (\ref{e3}) by $\frac{1}{2} (1 - \imath
\gamma_5)$.\\
However the Super Kamiokande experiment \cite{skexpt} clearly showed
that the neutrino has a small mass. On the other hand according to
the standard model the neutrino should be massless \cite{taylor}. So
in recent years there has been much work on going beyond the
standard model in order to explain amongst other things, the
neutrino mass. Currently, the dominant view is that the neutrino
mass oscillation arises from the MSW mechanism.\\
We would now like to deduce the neutrino mass from a slightly
different perspective, and argue that it is a result of a fuzzy
spacetime structure, resulting in a modified Dirac equation.
\section{Modified Dirac Equation}
Our starting point is the fact that, if there is a minimum
fundamental length $l$, the usual Quantum Mechanical commutation
relations get modified as shown by Snyder a long time ago
\cite{snyder,cu,uof,tduniv}. These relations are now replaced by
\begin{equation}
[x,p] = \hbar' = \hbar [1 + \left(\frac{l}{\hbar}\right)^2 p^2]\,
etc\label{5He2}
\end{equation}
(Cf. also ref.\cite{bgsust}). (\ref{5He2}) shows that effectively
$\hbar$ is replaced by $\hbar'$. So,
$$E = [m^2 + p^2 (1 + l^2 p^2)^{-2}]^{\frac{1}{2}}$$
or, the energy-momentum relation leading to the Klein-Gordon
Hamiltonian being modified to the so called Snyder-Sidharth
Hamiltonian \cite{glinka},
\begin{equation}
E^2 = m^2 + p^2 + \propto l^2 p^4,\label{5He3}
\end{equation}
neglecting higher order terms and using natural units, $c = 1 =
\hbar$ while $\propto$ is a dimensionless constant.\\
For Fermions the analysis can be more detailed, in terms of Wilson
lattices \cite{mont}. The free Hamiltonian now describes a
collection of harmonic fermionic oscillators in momentum space.
Assuming periodic boundary conditions in all three directions of a
cube of dimension $L^3$, the allowed momentum components are
\begin{equation}
{\bf q} \equiv \left\{q_k = \frac{2\pi}{L}v_k; k = 1,2,3 \right\},
\quad 0 \leq v_k \leq L - 1\label{4.59}
\end{equation}
(\ref{4.59}) finally leads to
\begin{equation}
E_{\bf q} = \pm \left(m^2 + \sum^{3}_{k=1} a^{-2} sin^2
q_k\right)^{1/2}\label{4.62}
\end{equation}
where $a = l$ is the length of the lattice, this being the desired
result. (\ref{4.62}) shows that $\alpha$ in (\ref{5He3}) is
positive. We have used the above analysis more to indicate that in
the Fermionic case, the sign of $\alpha$ is positive. A rigid
lattice structure imposes restrictions on the spacetime - for
example homogeneity and isotropy. Such restrictions are not demanded
by the author's model of fuzzy spacetime, and we use the lattice
model more as a computational device (Cf. ref.\cite{uof}). This
leads to a modification of the Dirac and Klein-Gordon equations at
ultra high energies (Cf.ref.\cite{uof,ijtp2,ijmpe2}). It may be
remarked that proposals like equation (\ref{5He3}) have been
considered by several authors based on phenomenological
considerations (Cf. refs.\cite{kif}-\cite{cam}). Our approach
however, has been fundamental rather than phenomenological. In any
case, modified Hamiltonians like (\ref{5He3}) have been considered
by other authors like Glashow and Coleman, though from a purely
phenomenological angle.\\
Once we consider a discrete spacetime structure, the energy momentum
relation, gets modified \cite{cu,mont} and we have,
\begin{equation}
E^2 - p^2 - m^2 - \propto l^2 p^4 = 0\label{6ce1}
\end{equation}
$l$ being the minimum length interval, which could be the Planck
length or more generally the Compton length. Let us now consider the
Dirac equation
\begin{equation}
\left\{ \gamma^\mu p_\mu - m\right\} \psi \equiv \left\{\gamma^\circ
p^\circ + \Gamma \right\} \psi = 0\label{6ce2}
\end{equation}
If we include the extra effect shown in (\ref{6ce1}) we get
\begin{equation}
\left(\gamma^\circ p^\circ + \Gamma + \beta l p^2\right) \psi =
0\label{6ce3}
\end{equation}
$\beta$ being a suitable matrix.\\
Multiplying (\ref{6ce3}) by the operator
$$\left(\gamma^\circ p^\circ - \Gamma - \beta l p^2\right)$$
on the left we get
\begin{equation}
p^2_0 - \left(\Gamma \Gamma + \left\{\Gamma \beta + \beta
\Gamma\right\} + \beta^2 l^2 p^4\right\} \psi = 0\label{6ce4}
\end{equation}
If (\ref{6ce4}), as in the usual theory, has to represent
(\ref{6ce1}), then we require that the matrix $\beta$ satisfy
\begin{equation}
\Gamma \beta + \beta \Gamma = 0, \quad \beta^2 = 1\label{6ce5}
\end{equation}
In $\Gamma$ of (\ref{6ce5}), we can at ultra high energies neglect
the mass $m$, as is well known in the theory of the Cini-Tauschek
transformation (Cf.ref.\cite{schweber}). (Cf.Section 4 remark 4).\\
From the properties of the Dirac matrices \cite{bd} it then follows
that (\ref{6ce5}) is satisfied if
\begin{equation}
\beta = \gamma^5\label{6ce6}
\end{equation}
Using (\ref{6ce6}) in (\ref{6ce3}), the modified Dirac equation
finally becomes, to a good approximation
\begin{equation}
\left\{\gamma^\circ p^\circ + \Gamma + \imath \alpha \gamma^5 l
p^2\right\} \psi = 0\label{6ce7}
\end{equation}
owing to the fact that we have \cite{bd}
\begin{equation}
P \gamma^5 = -\gamma^5 P\label{6ce8}
\end{equation}
It follows that the so called Dirac-Sidharth equation (\ref{6ce7})
is not invariant under reflections. This is a result which is to be
expected because the correction to the usual energy momentum
relation, as shown in (\ref{6ce1}) arises when $l$ is of the order
of the Compton wavelength. The usual Dirac four spinor
$\left(\begin{array}{ll}\phi \\ \chi
\end{array}
\right)$ has the so called positive energy (or large) components
$\phi$ and the negative energy (or small) components $\chi$.
However, when we approach the Compton wavelength, that is as
$$p \to mc$$
the roles are reversed and it is the $\chi$ components which
predominate. Moreover the $\chi$ two spinor behaves under reflection
as \cite{bd}
$$\chi \to -\chi$$
In any case, this too provides an experimental test. We can also see
that due to the modified Dirac equation (\ref{6ce7}), there is no
additional effect on the anomalous gyromagnetic ratio. This is
because, in the usual equation from which the magnetic moment is
determined \cite{merz} viz.,
$$\frac{d\vec{S}}{dt} = -\frac{e}{\mu c} \vec{B} \times \vec{S},$$
where $\vec{S} = \hbar \sum/2$ is the electron spin operator, there
is now an extra term
\begin{equation}
\left[\gamma^5, \sum\right]\label{6ce9}
\end{equation}
However the expression (\ref{6ce9}) vanishes by the property of the
Dirac matrices.\\
\section{Massive Neutrinos}
Taking units $c = 1 = \hbar$ again, the equation (\ref{6ce7}) can be
written as
\begin{equation}
- \gamma_0 p^0 \psi = (D + \imath \alpha lp^2 \gamma_5)
\psi\label{A}
\end{equation}
In (\ref{A}) $D$ represents the usual Dirac operator given in
(\ref{6ce7}) and there is the extra term following it. Equation
(\ref{A}) is valid approximately for a massive and exactly for a
massless Dirac particle \cite{bgs}. We can see that, as the
Hamiltonian is given by (Cf. Section 1).
\begin{equation}
H = \imath \gamma_5 \vec{\sum} \cdot \vec{p} = \imath \gamma_5
|\vec{p}| s (\vec{p})|\label{B}
\end{equation}
the extra term in (\ref{A}) represents a mass term. In other words
due to the Hamiltonian (\ref{5He3}), the Dirac particle acquires an
additional mass. However what is very interesting is that the extra
term is not invariant under parity owing to the presence of
$\gamma_5$. Indeed as we know from the theory of Dirac matrices
\cite{bd}
\begin{equation}
P \gamma_5 = - P \gamma_5\label{C}
\end{equation}
Let us now consider the case of a massless Dirac particle. We can
see that in this case (\ref{A}) can represent the neutrino with a
mass, there now being no need for the extra constraint (\ref{e9})
required for massless neutrinos. Equation (\ref{A}) automatically
gives a parity non conserving particle. In other words a massless
particle, satisfying the Dirac equation in the usual theory now
acquires a mass.
\section{Remarks}
1. We can now ask what does (\ref{A}) represent if to start with the
particle has a mass $m$? As can be seen it now acquires an extra
mass, at ultra high energies. However this extra mass does not
conserve parity. So, usual particles at very high energies become
unstable, due to this additional contribution: the mass now has two
parts, the usual mass $m$ that is invariant under reflection, but
also a parity non-conserving part. We could also think of it as
follows: a usual spin half particle of mass $m$, at ultra high
energies shows up as two other particles with slightly different
masses. It appears that this could be a case like
the $K_0$ meson.\\
2. The standard model or the Dirac equation are strictly in
accordance with Special Relativity and the neutrino mass is
accordingly zero. Now however, we have the Hamiltonian (\ref{5He3})
which is a very high energy correction to the usual relativistic
dispersion relation. It is this modification or extra term that
throws up the
massive neutrino breaking the Lorentz symmetry.\\
3. We can look upon this in a different way \cite{heine}. The usual
Dirac equation which is invariant under the Lorentz transformation
including the improper parity operation is given by the
representation
\begin{equation}
D^{(\frac{1}{2}0)} \oplus D^{(0 \frac{1}{2})}\label{D}
\end{equation}
The solutions which are according to the two-component
representation $D^{(\frac{1}{2} 0)}$ or $D^{(0 \frac{1}{2})}$ are
not invariant under parity. However the combined four-component
solution Dirac spinor in (\ref{D}) is invariant under a Lorentz
transformation plus the parity transformation. When we introduce the
extra term in the modified Dirac equation, this term spoils the
invariance under parity. If we write the usual Dirac spinor as
\begin{equation}
\psi = \left(\begin{array}{ll}\phi \\\chi\end{array}\right)\label{E}
\end{equation}
then it is known as we saw \cite{bd2} that at very high energies the
upper or positive two spinor $\phi$ is invariant under parity, but
not the lower. That is under the parity operator ${\bf P}$
\begin{equation}
\phi \to \phi , \chi \to - \chi\label{F}
\end{equation}
Now $\phi$ and $\chi$ each are given by the $D^{(\frac{1}{2}0)}$ and
$D^{(0\frac{1}{2})}$ representations. Under space reflections, they
go into each other and it is for this reason that $\psi$ given by
(\ref{D}) is invariant.\\
If we still consider this solution as an approximation to (\ref{A})
also, the result of the parity operator ${\bf P}$ would now be to
interchange the behavior $\phi$ and $\chi$ under ${\bf P}$.\\
\section{Appendix}
I. It is well known that at ultra high energies the massive Dirac
equation goes over the massless Dirac equation because of the
dominance of the kinetic energy term. In this case there is the well
known Cini transformation which reduces the Dirac Hamiltonian to the
form
$$H = \frac{\vec{\propto} \cdot \vec{p}}{|\vec{p}|} E (\vec{p})$$
where
\begin{equation}
E(\vec{p}) = c \left[\vec{p}^2_0 +
m^2c^2\right]^{\frac{1}{2}}\label{eb}
\end{equation}
The Dirac equation now throws up the positive and negative helicity
states which are described by two separate two component equations.
The implication of the extra term in the Hamiltonian (\ref{5He3})
can be seen from (\ref{eb}). For example if we take the massless
case $m = 0$, there is now a new mass term in (\ref{eb}). However,
because of the presence of $\gamma^5$ in the extra term, the
helicity states now have two different masses indicating that the
righthanded anti neutrino would have a slightly different mass
compared to the lefthanded neutrino.\\
II. It may be mentioned that starting from 1985, there were attempts
to introduce a tachyonic neutrino by introducing ad hoc an equation
resembling (\ref{A}), but over the years these efforts have not led
to anything tangible.\\
III. It is well known that the standard model of particle physics is
as of now the most complete theory and yet there are frantic efforts
to go beyond the standard model to overcome its shortcomings. Some
of
these are:\\
1. It fails to deliver the mass to the neutrino which thus remains a
massless particle in this theory.\\
2. This apart it does not include gravity, which is otherwise one of
the four fundamental interactions.\\
3. There is the hierarchy problem viz., the wide range of masses for
the elementary particles or even for the quarks.\\
4. It appears that other as yet undiscovered particles exist which
could change the picture, for example in supersymmetry in which the
particles have their supersymmetric counterparts.\\
5. The standard model has no place for dark matter, which on the
other hand has not yet been definitely found. Nor is there place for dark energy.\\
6. Finally one has to explain the $18$ odd arbitrary constants which
creep into the theory.\\
There are however obvious shortcomings which can be addressed in a
relatively simple manner which could enable us to go beyond the
standard model. Let us start with the standard model Lagrangian
$$LGWS = \sum_{f} (\bar{\Psi}_f (\imath \gamma^\mu \partial_\mu -
m_f) \Psi_f - eQ_f \Psi_f \gamma^\mu \Psi_f A_\mu) +$$
$$+ \frac{g}{\sqrt{2}} \sum_{\imath} (\bar{a}^{\imath}_{L}
\gamma^\mu b^\imath_L W^+_\mu + \bar{b}^\imath_L \gamma^\mu
a^\imath_L W^-_\mu ) + \frac{g}{2C_w} \sum_f \bar{\Psi}_f \gamma^\mu
(I^3_f - 2S^2_w Q_f - I^3_f \gamma_5) \Psi_f Z_\mu +$$
$$-\frac{1}{4} |\partial_\mu A_\nu - \partial_\nu A_\mu - \imath e
(W^-_\mu W^+_\nu - W^+_\mu W^-_\nu)|^2 - \frac{1}{2}| \partial_\mu
W^+_\nu - \partial_\nu W^+_\mu +$$
$$-\imath e (W^+_\mu + A_\nu - W^+_\nu A_\mu ) + \imath g' c_w
(W^+_\mu Z_\nu - W^+_\nu Z_\mu|^2 +$$
$$- \frac{1}{4}|\partial_\mu Z_\nu - \partial_\nu Z_\mu + \imath g'
c_w (W^-_\mu W^+_\nu - W^+_\mu W^-_\nu)|^2 +$$
$$-\frac{1}{2}M^2_\eta \eta^2 - \frac{gM^2_\eta}{8M_W} \eta^3 -
\frac{g^{'2}M^2_\eta}{32M_W} \eta^4 + |M_W W^+_\mu + \frac{g}{2}
\eta W^+_\mu|^2 +$$
\begin{equation}
+\frac{1}{2} |\partial_\mu \eta + \imath M_Z Z_\mu + \frac{\imath
g}{2C_w} \eta Z_\mu |^2 - \sum_{f} \frac{g}{2} \frac{m_F}{M_W}
\bar{\Psi}_f \Psi_f \eta\label{e1}
\end{equation}
which includes the Dirac Lagrangian amongst other things.\\
We would now like to point out that all this has been on the basis
of the usual point spacetime which is what may be called
commutative. But in recent years several authors including in
particular the present author have worked with a noncommutative
spacetime which originates
back to Snyder in the late forties itself. (This was in an attempt to overcome the divergences).\\
We first observe that it was Dirac \cite{dirac} who pointed out two
intriguing features of his equation: 1. The Compton wavelength and
2. Zitterbewegung.\\
For the former, his intuition was that we can never make
measurements at space or time points. We need to observe over an
interval to get a meaningful definition of momentum for example.
This interval was the Compton region \cite{bgsdas}. Next, his
solution was rapidly oscillatory, what is called Zitterbewegung.
This oscillatory behaviour disappears on averaging over spacetime
intervals over the Compton region. Once this is done while
meaningful physics appears, we are left with not points but minimum
intervals.\\
This leads to a noncommutative geometry. One model for this is that
of Snyder \cite{snyder}. Applied at the Compton wavelength this
leads to the so called Snyder-Sidharth dispersion relation, the
geometry being given by \cite{tduniv}
\begin{equation}
[x_\imath , x_j] = \beta_{\imath j} \cdot l^2\label{6eeA}
\end{equation}
As described in details in reference \cite{bgsijmpe2010} this leads
to a modification in the Dirac and also the Klein-Gordon equation.
This is because (\ref{6eeA}) in particular it leads to the following
energy momentum relation (Cf.ref.\cite{tduniv})
\begin{equation}
E^2 - p^2 - m^2 + \alpha l^2 p^4 = 0\label{6ce1}
\end{equation}
where $\alpha$ is a scalar constant, $\sim 10^{-3}$
\cite{bgsdasarka,bgsdasarka2}. Though the value of $\alpha$ is of no
consequence for the present work, it may be mentioned that $\alpha$
gives the Schwinger term. If we work with this energy momentum
relation (\ref{6ce1}) and follow the usual process we get as in the
usual Dirac theory
\begin{equation}
\left\{ \gamma^\mu p_\mu - m\right\} \psi \equiv \left\{\gamma^\circ
p^\circ + \Gamma \right\} \psi = 0\label{6ce2}
\end{equation}
We now include the extra term in the energy momentum relation
(\ref{6ce1}). It can be easily shown that this leads to
\begin{equation}
p^2_0 - \left(\Gamma \Gamma + \left\{\Gamma \beta + \beta
\Gamma\right\} + \beta^2 \alpha l^2 p^4\right\} \psi = 0\label{6ce4}
\end{equation}
Whence the modified Dirac equation
\begin{equation}
\left\{\gamma^\circ p^\circ + \Gamma + \gamma^5 \alpha l p^2\right\}
\psi = 0\label{6ce7}
\end{equation}
The Modified Dirac equation contains an extra term. The extra term
gives a slight mass for the neutrino which is roughly of the correct
order viz., $10^{-8} m_e$, $m_e$ being the mass of
the electron. The behaviour too is that of the neutrino \cite{bgsijmpe2010,nap11}.\\
To sum up the introduction of the noncommutative geometry described
in (\ref{6eeA}) leads to a Dirac like equation (\ref{6ce7}) and a
Lagrangian that leads to the
neutrino mass at least the electron neutrino to start with.\\
It must be pointed out that the modified Lagrangian differs from the
usual Lagrangian in that the $\gamma^0$ matrix is now replaced by a
new matrix
$$\gamma^{0'} = \gamma^0 + \gamma^0 \cdot \gamma^5 lp^2$$
that includes the term giving rise to the neutrino mass. We can
verify that the modified Lagrangian gives back the modified Dirac
equation (\ref{6ce7}). Further as has been discussed in detail the
extra term arising out of the noncommutative geometry is the direct
result of the dark energy which thus also features in the modified
standard model Lagrangian. This apart this argument has been shown
to lead to a mass spectrum for elementary particles that includes
all the elementary particles, giving the masses with about $5 \%$ or
less error \cite{tduniv}.


\begin{thebibliography}{99}
\bibitem {schweber} S.S. Schweber. (1961). \emph{An Introduction to Relativistic Quantum
Field Theory} (Harper and Row, New York), p.108ff.
\bibitem {skexpt} SuperKamiokande Website \emph{http://www-sk.icrr.u-tokyo.ac.jp/}
\bibitem {taylor} Greiner, W. and Reinhardt, J. (1990). \emph{Gauge Theories of Weak Interactions} (Springer-Verlag, Berlin).
\bibitem {snyder} Snyder, H.S. (1947). \emph{Physical Review} Vol.72, No.1, July 1 1947, pp.68--71.
\bibitem {cu} Sidharth, B.G. (2001). \emph{Chaotic Universe: From the Planck to the Hubble Scale}
(Nova Science, New York).
\bibitem {uof} Sidharth, B.G. (2005). \emph{The Universe of Fluctuations} (Springer,
Netherlands).
\bibitem {tduniv} Sidharth, B.G. (2008). \emph{The Thermodynamic
Universe} (World Scientific, Singapore).
\bibitem {bgsust} Sidharth, B.G. (2008). \emph{Foundations of Physics} 38 (1), pp.89-96.
\bibitem {glinka} Glinka, L.A. \emph{arXive. 0812.0551}; \emph{Apeiron} April 2009.
\bibitem {mont} Montvay, I. and Munster, G. (1994). \emph{Quantum Fields on a
Lattice} (Cambridge University Press) pp.174ff.
\bibitem {ijtp2} Sidharth, B.G. (2004). \emph{Int.J.of Th.Phys.} Vol.43, (9), September 2004,
pp.1857-1861.
\bibitem {ijmpe2} Sidharth, B.G. (2005). \emph{Int.J.Mod.Phys.E.} 14, (6), pp.923ff.
\bibitem {kif} Kifune, T. (1999). \emph{astro-ph/9904164}; \emph{Astrophys. J. Lett.}
518, pp.L21.
\bibitem {pro} Protheroe, R.J. and Meyer, H. (2000). \emph{Phys.Lett.}
B493, pp.1.
\bibitem {alo} Aloisio, R., Blasi, P., Ghia, P.L. and Grillo,
A.F. (2000). \emph{Phys.Rev.} D62, pp.053010.
\bibitem {kluz} Kluzniak, W. \emph{astro-ph/9905308}.
\bibitem {sato} Sato, H. \emph{astro-ph/0005218}.
\bibitem {amel2} Amelino-Camelia, G. and Piran, T. (2001). \emph{Physics Letters}
B497, pp.265--270.
\bibitem {amel3} Amelino-Camelia, G. \emph{gr-qc/0012051v2} (He proposes a
conceptual framework in which deformed dispersion relations coexist
with a relativistic description of the short distance structure of
spacetime).
\bibitem {amel4} Amelino-Camelia, G. and Piran, T. \emph{Phys.Rev. D.}
Vol.64, pp.036005.
\bibitem {amel5} Amelino-Camelia, G., John Ellis, Mavnomatos, N.E.,
Nanopoulos, D.V. and Subir Sarkar. (1998). \emph{Nature} 393, 25
June, 1998, pp.763-765; \emph{(astro-ph/9712103 v2 17 April 1998)}.
\bibitem {cam} Amelino-Camelia, G. (2002). \emph{Nature} Vol.418, 4 July 2002.
\bibitem {bd} Bjorken, J.D. and Drell, S.D. (1964). \emph{Relativistic Quantum Mechanics}
(Mc-Graw Hill, New York), pp.39.
\bibitem {merz} E. Merzbacher. (1970). \emph{Quantum Mechanics} (Wiley, New
York); Greiner, W. (1983). \emph{Relativistic Quantum Mechanics:
Wave Equation} 2nd Ed. (Springer);  Greiner, W. and Reinhardt, J.
(1987). \emph{Quantum Electrodynamics} 3rd Ed. (Springer).
\bibitem {bgs} Sidharth, B.G. \emph{arXiv. 0811.4541}, to appear in
\emph{Int.J.Mod.Phys.E.}
\bibitem {heine} Heine, V. (1960). \emph{Group Theory in Quantum Mechanics} (Pergamon Press,
Oxford), pp.364.
\bibitem {bd2} Bjorken, J.D. and Drell, S.D. (1965). \emph{Relativistic Quantum Fields}
(McGraw-Hill Inc., New York), pp.44ff.
\bibitem {dirac} Dirac, P.A.M. (1958). \emph{The Principles of
Quantum Mechanics} (Clarendon Press, Oxford), pp.4ff, pp.253ff.
\bibitem {bgsdas} Sidharth, B.G. and Abhishek, Das. (2017).
\emph{Int.J.Mod.Phys.A.} Vol.32, 2017, pp.1750173ff.
\bibitem {bgsijmpe2010} Sidharth, B.G. (2010). \emph{Int.J.Mod.Phys.E}19(11),2010, pp.1-8.
\bibitem {bgsdasarka} Sidharth, B.G., Das, A., Arka, R. (2015). \emph{Electron J. Theor. Phys.} 12, No.34, 2015, pp.139-152.
\bibitem {bgsdasarka2} Sidharth, B.G., Das, A., Arka, R. (2016). \emph{Int.J.Th.Phys.} Volume 55, Issue 2, 2016, pp. 801-808.
\bibitem {nap11} Sidharth, B.G. (2017). \emph{New Advances in Physics} 11 (2) 2017, pp.5-97.
\end{thebibliography}
\end{document}